\documentclass[apj]{emulateapj}
\usepackage{apjfonts}
\submitted{Received 2010 February 27; accepted 2010 June 15; published --}
\journalinfo{Accepted to ApJ, 2010 June 15}

\newcommand{\FeI}{\ion{Fe}{1}}
\newcommand{\CaII}{\ion{Ca}{2}}

\newcommand{\HeI}{\ion{He}{1}}

\newcommand{\NaI}{\ion{Na}{1}}
\newcommand{\MgI}{\ion{Mg}{1}}
\newcommand{\kms}{km$\,$s$^{-1}$}
\newcommand{\arcpix}{$^{\prime\prime}$~pixel$^{-1}$}
\newcommand{\ha}{H$\alpha$}
\newcommand{\hb}{H$\beta$}

\shorttitle{DOPPLER SHIFT AND ASYMMETRY OF STOKES PROFILES OF \FeI\ AND \MgI\ LINES}
\shortauthors{DENG ET AL.}

\begin{document}
\title{On the Doppler Shift and Asymmetry of Stokes Profiles of Photospheric \FeI\ and Chromospheric \MgI\ Lines}
\author{Na Deng and Debi Prasad Choudhary}
\affil{California State University Northridge, Physics
       and Astronomy Department, 18111 Nordhoff St., Northridge, CA 91330; na.deng@csun.edu, debiprasad.choudhary@csun.edu}
\author{ K. S. Balasubramaniam}
\affil{USAF/Air Force Research Laboratory, Solar Disturbances Prediction ,
       P.O.~Box 62, Sunspot, NM 88349, U.S.A.; bala@nso.edu}

\begin{abstract}
We analyzed the full Stokes spectra using simultaneous measurements of the photospheric (\FeI\ 630.15 and 630.25~nm) and chromospheric (\MgI\ b$_2$ 517.27~nm) lines. The data were obtained with the HAO/NSO Advanced Stokes Polarimeter, about a near disc center sunspot region, NOAA AR 9661. We compare the characteristics of Stokes profiles in terms of Doppler shifts and asymmetries among the three spectral lines, which helps us to better understand the chromospheric lines and the magnetic and flow fields in different magnetic regions. The main results are: (1) For penumbral area observed by the photospheric \FeI\ lines, Doppler velocities derived from Stokes $I$ ($\nu_i$) are very close to those derived from linear polarization profiles ($\nu_{lp}$) but significantly different from those derived from Stokes $V$ profiles ($\nu_{zc}$), which provides direct and strong evidence that the penumbral Evershed flows are magnetized and mainly carried by the horizontal magnetic component. (2) The rudimentary inverse Evershed effect observed by the \MgI\ b$_2$ line provides a qualitative evidence on its formation height that is around or just above the temperature minimum region. (3) $\nu_{zc}$ and $\nu_{lp}$ in penumbrae and $\nu_{zc}$ in pores generally approach their $\nu_i$ observed by the chromospheric \MgI\ line, which is not the case for the photospheric \FeI\ lines. (4) Outer penumbrae and pores show similar behavior of the Stokes $V$ asymmetries that tend to change from positive values in the photosphere (\FeI\ lines) to negative values in the low chromosphere (\MgI\ line). (5) The Stokes $V$ profiles in plage regions are highly asymmetric in the photosphere and more symmetric in the low chromosphere. (6) Strong red shifts and large asymmetries are found around the magnetic polarity inversion line within the common penumbra of the $\delta$ spot. We offer explanations or speculations to the observed discrepancies between the photospheric and chromospheric lines in terms of the three-dimensional structure of the magnetic and velocity fields. This study thus emphasizes the importance of spectro-polarimetry using chromospheric lines.

\end{abstract}


\section{INTRODUCTION}
Solar photospheric magnetic fields and their associated dynamics have been extensively studied. However an understanding of their vertical variation from the photosphere though the chromosphere to the corona needs better comprehension. Direct measurement and diagnosis of magnetic and flow fields in higher layers, especially in the near force-free chromosphere can play an important role in fully understanding their three-dimensional (3D) structures. Nevertheless, such efforts were relatively rare due to the paucity of chromospheric spectral lines with suitable Zeeman-split sensitivity \citep{Dalgarno+Layzer1987soap.conf.....D} and the complicated dynamics and topology of the magnetized plasma in this particular layer. Taking advantage of modern observational techniques and recognizing the critical need to explore chromospheric magnetic fields, increasing attention has recently been paid to the spectro-polarimetry of the chromosphere \citep[e.g.,][]{Socas-Navarro+etal2000Sci...288.1396S, Zhang+Zhang2000SoPh..194...19Z, TrujilloBueno+etal2002Natur.415..403T, LopezAriste+Casini2002ApJ...575..529L, Solanki+etal2003Natur.425..692S, leka+metcalf2003, Lagg+etal2004A&A...414.1109L, Balasubramaniam+etal2004ApJ...606.1233B}, which reveal the height-dependent variation of different types of magnetic features. Those analyses were made using magnetically sensitive spectral lines formed in the chromospheric levels, such as \CaII\ 849.8 and 854.2~nm, \HeI\ 1083.0~nm and D$_3$ line, \NaI\ D$_1$ 589.6~nm, \ha\ and \hb\ lines. In particular, the \MgI\ b$_2$ 517.27~nm line is favorable for spectro-polarimetry in the low chromosphere because its core forms in a narrow region near or right above the temperature minimum region (TMR) with a relatively large Land\'{e} g$_{eff}$ factor of 1.75 \citep[e.g.,][]{Briand+Solanki1995A&A...299..596B}. Theoretical calculation \citep{Altrock+Canfield1974, Lites+etal1988, Mauas+etal1988} and observational analysis \citep{Briand+Solanki1998A&A...330.1160B, Briand+MartinezPillet2001ASPC..236..565B, Gosain+Choudhary2003SoPh..217..119G} have been carried out on \MgI\ b$_2$ Stokes polarization spectra, which corroborate that it is well suited for probing the magnetic and flow fields in the low chromosphere.

Doppler shifts of Stokes $I$ and $Q, U, V$ signals can be used to investigate mass flows in and around magnetic elements \citep{Solanki1986A&A...168..311S,
Solanki+Pahlke1988A&A...201..143S}. In contrast to the Stokes $I$ that represents the total intensity of light, the Stokes $Q, U$ (the linearly polarized component) and $V$ (the circularly polarized component) have contribution only from the magnetized plasma. Thus Doppler shifts of the Stokes $I$ and $Q, U, V$ profiles reflect average line-of-sight (LOS) velocities of plasmas in the whole resolution element and those within magnetic regions, respectively \citep[e.g.,][]{Solanki1986A&A...168..311S}.

Asymmetries of Stokes profiles are a powerful diagnostics of the height and spatial properties of magnetic and flow fields \citep[e.g.,][]{Grigorev+Katz1975SoPh...42...21G}. Under a restrictive Milne-Eddington atmosphere where magnetohydrodynamic (MHD) conditions (e.g., plasma velocity, magnetic field vector, and temperature) are constant with depth and the source function varies linearly with optical depth \citep{Unno1956, Landi+Landi1985}, the Zeeman-split Stokes $I,Q,U$ profiles are symmetric, while $V$ is antisymmetric about the line center. Most of the observed Stokes spectra, however, deviate from the ideal forms and exhibit notable difference between the blue and red wings in amplitude and/or area. Therefore the Stokes asymmetry is able to reveal the inhomogeneous nature of the solar atmosphere \citep[e.g.,][]{illing+landmann+mickey1974A&A....35..327I, illing+landmann+mickey1974A&A....37...97I}. Direct observations \citep[e.g.,][]{Balasubramaniam+etal1997ApJ...482.1065B, Sigwarth2001} and modeling of synthetic Stokes profiles \citep[e.g.,][]{Solanki+Pahlke1988A&A...201..143S, SanchezAlmeida+Lites1992ApJ...398..359S, Solanki+Montavon1993A&A...275..283S, SanchezAlmeida+etal1996ApJ...466..537S, Grossmann-Doerth+etal2000, Steiner2000SoPh..196..245S} have found possible physical conditions for producing asymmetric Stokes spectra, such as gradients of flow velocity and/or magnetic field vector along the LOS direction and nonuniform atmosphere that contains two or more distinct magnetic and/or flow components in one resolution element. A more comprehensive study by \citet{LopezAriste2002ApJ...564..379L} based on the analytical solution for the generalized transfer equation explicitly demonstrates that the velocity gradients are the necessary and sufficient condition for Stokes profile asymmetries. The addition of other effects or gradients of other MHD parameters (e.g., magnetic field and temperature) could enhance the asymmetries and give different characteristics of asymmetries. For example, strong Stokes $I$ and $V$ asymmetries are usually found around magnetic flux concentrations in the photosphere, which is attributed to the presence of the ``magnetic canopy'' \citep{grossmanndoerth+schuessler+solanki1988A&A...206L..37G, grossmanndoerth+schuessler+solanki1989A&A...221..338G, Solanki1989A&A...224..225S, leka+steiner2001}. Strong gradients in both magnetic and flow fields are naturally generated because the ``magnetic canopy'' spatially separates the field-free convective downdrafts from the more static flux tube that fans out above the photosphere.

The aforementioned analyses of Stokes profiles predominantly use photospheric spectral lines. In retrospect, only few such attempts using chromospheric Stokes spectra have been reported. As examples, \citet{Briand+Solanki1998A&A...330.1160B} inferred fast down flows concentrated in narrow lanes surrounding the flux tube below the ``magnetic canopy'' by using Stokes $V$ asymmetries of \MgI\ b$_2$ line in additional to photospheric lines. \citet{Gosain+Choudhary2003SoPh..217..119G} compared the Stokes $V$ asymmetries between photospheric \FeI\ lines and chromospheric \MgI\ lines and found differences in their spatial distribution within a simple sunspot. Additional investigation of chromospheric Stokes spectra in conjunction with photospheric measurements would improve our understanding of 3D magnetic topology and plasma flows.

\begin{figure}[t]
  \epsscale{1.17}
  \plotone{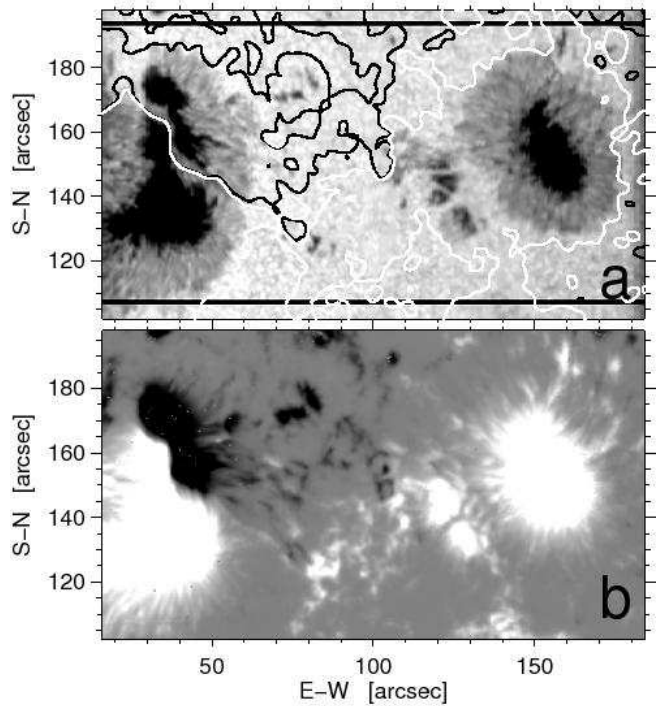}
  \caption{(\textit{a}) The \FeI\ 630.15~nm continuum image of the NOAA AR 9661 on 2001 October~17. White and black contours outline the positive and negative polarities, respectively. (\textit{b}) The corresponding longitudinal magnetic field mimicked by the peak value of Stokes $V$ blue lobe.}
  \label{FIG:AR}
\end{figure}

In this paper, we study Doppler shifts and asymmetries of Stokes profiles of three spectral lines simultaneously observed in an active region close to the disk center. The three spectral lines, \FeI\ 630.25~nm (g$_{eff} = 2.5$), \FeI\ 630.15~nm (g$_{eff} = 1.67$), and \MgI\ b$_2$, are formed at partially overlapping atmospheric layers spanning from the photosphere to the low chromosphere. Even though both are regarded as photospheric lines, the \FeI\ 630.15~nm line often forms above the 630.25~nm line \citep[e.g.,][]{Khomenko+Collados2007ApJ...659.1726K}. Combining data from multiple atmospheric heights using these spectral lines thus enables us to shed new light on the properties of magnetic and flow structures associated with different magnetic features. The plan of this paper is as follows: in \S\S~\ref{sec:observation} and~\ref{sec:reduction}, we introduce the observations and present the procedure for data reduction. In \S~\ref{sec:results} the results of comparison of profile shifts and asymmetries among the three spectral lines are described and discussed. We summarize our major findings and present some prospects for future studies in \S~\ref{sec:summary}.

\section{OBSERVATIONS}\label{sec:observation}
We obtained the spectropolarimetric data sets using the 76~cm Dunn Solar Telescope and its associated Adaptive Optics system \citep{rimmele2004a}, along with the High Altitude Observatory/National Solar Observatory Advanced Stokes Polarimeter \citep[HAO/NSO ASP,][]{Elmore+etal1992, Lites1996, skumanich+etal1997}. The observations were obtained on 2001 October 17 under excellent seeing conditions. The ASP is a slit-grating spectropolarimeter. The 1.6$^{\prime}$~$\times$~0.6\arcsec\ slit was oriented in the solar north-south direction and scanned in the east-west direction. Each whole scan took 280 steps with a step size of 0.6\arcsec, which results in a field of view (FOV) of 1.6$^{\prime}$~$\times$~2.8$^{\prime}$ that covers the near disk center (N14$^\circ$, W6$^\circ$, $\mu= $cos$\theta=0.96$) active region NOAA 9661 as shown in Figure~\ref{FIG:AR}. It contains a $\delta$ sunspot at left, an $\alpha$ sunspot at right, as well as several pores and plages lying in between. The obtained spatial sampling is about 0.42\arcpix\ in the north-south direction and 0.6\arcpix\ in the east-west direction. Full Stokes $I,Q,U,V$ spectra were taken simultaneously in two spectral bands centered at 630.2~nm and 517.27~nm by two ASP cameras. The dispersions for the two spectral bands are 1.27 and 1.02~pm~pixel$^{-1}$, respectively. Each whole scan took about 25~minutes. A total of five whole scans were continuously taken within 2.5~hours, during which no significant evolution of the magnetic structures was found. Other information about the observation run can be found in \citet{Choudhary+Bala2007}.

\section{DATA REDUCTION}\label{sec:reduction}
We applied the standard ASP calibration procedures \citep{Lites+etal1991sopo.work....3L, skumanich+etal1997} to the data sets. The calibrated Stokes $I, Q, U, V$ spectra were normalized to the quiet Sun continuum intensity $I_{c}$ that was obtained by fitting the Kitt Peak FTS atlas to the observed Stokes $I$ profiles. The data sets from the five whole scans were then integrated to increase the signal to noise ratio (S/N). Furthermore, we calculated the standard deviation $\sigma$ in near continuum wavelength range for each integrated Stokes $I, Q, U, V$ profile to represent the profile noise level. Only profiles with signal amplitude greater than 7$\sigma$ (an empirical S/N threshold that performs well in our data analysis) were used for this study.

We extracted the parameters for analyzing the shift and asymmetry of Stokes profiles using the following procedures.

(1) \emph{Stokes $I$ LOS velocities $\nu_i$}. We employed a Fourier phase method \citep[see][]{schmidt+stix+woehl1999} to determine the line core wavelength. We then converted the wavelength shifts to velocities ($\nu_i$) by using the Doppler formula. As a frame of reference, we use the average velocity of a small area in the nearly motionless umbra.

(2) \emph{Stokes $V$ LOS velocities $\nu_{zc}$}. Normalized Stokes $V$ profiles were smoothed to remove local noise. They were then divided into two cases: normal profiles that have two opposite lobes and one zero-crossing (ZC) point in between, and abnormal profiles that have no or multiple ZCs within a wavelength range corresponding to $\pm 5$~\kms\ from Stokes $I$ line center. Using this method, there is a small chance that we may improperly classify normal profiles into an abnormal class. For example, strong magneto-optical effects (see Fig.~\ref{FIG:ABVSample}$c$ and text in \S~\ref{abnormal}) produce Stokes $V$ profiles with 3 ZCs, although the profile is essentially normal. Such improperly classified cases only account for a small portion ($<$ 0.3\%) in the data and do not affect the analysis. For normal profiles, the ZC shifts are converted to LOS velocities ($\nu_{zc}$) in a same manner as $\nu_i$. The abnormal profiles will be discussed in \S~\ref{abnormal}.

(3) \emph{Linear polarization (LP) LOS velocities $\nu_{lp}$}. The combination $(Q^2+U^2)^{1/2}$ is a measure of the overall $LP$ magnitude \citep[e.g.,][]{leka+steiner2001}. The $LP$ profiles were smoothed to remove local noise. It is difficult to measure the position of the central $\pi$ component as it is usually small and sometimes even vanishes (see Figure~\ref{FIG:AVGPROF}). Instead we measure the peak positions of the two $\sigma$ components and use their mid-point as the position of the $LP$ profile. We then converted the $LP$ profile shifts to LOS velocities ($\nu_{lp}$) in a same manner as $\nu_i$.

(4) \emph{Amplitude asymmetry $\delta a$ and area asymmetry $\delta A$}. The peak amplitudes of the blue ($a_b$) and the red ($a_r$) lobes of normal Stokes $V$ profiles and those of the two $\sigma$ components of $LP$ profiles are determined. The areas of blue ($A_b$) and red ($A_r$) lobes are obtained by integrating over the same wavelength range (0.02~nm) on both sides of the ZC wavelength for Stokes $V$ profiles or the mid-point of the two $\sigma$ components for $LP$ profiles. The integrating range not only covers the majority of the Stokes $V$ or $LP$ signal corresponding to Stokes $I$ Doppler core, but also rules out the blends from the neighboring telluric lines. Following \citet{Solanki+Stenflo1984A&A...140..185}, the amplitude asymmetry $\delta a$ and area asymmetry $\delta A$ are derived using the following formulae:
\begin{equation}
   \delta a = \frac{|a_b| - |a_r|}{|a_b| + |a_r|}  ~,
\label{eq:ampasym}
\end{equation}

\begin{equation}
   \delta A = \frac{|A_b| - |A_r|}{|A_b| + |A_r|}  ~.
\label{eq:areasym}
\end{equation}\\

\section{RESULTS AND DISCUSSIONS}\label{sec:results}
\begin{figure*}[t]
  \epsscale{1.17}
  \plotone{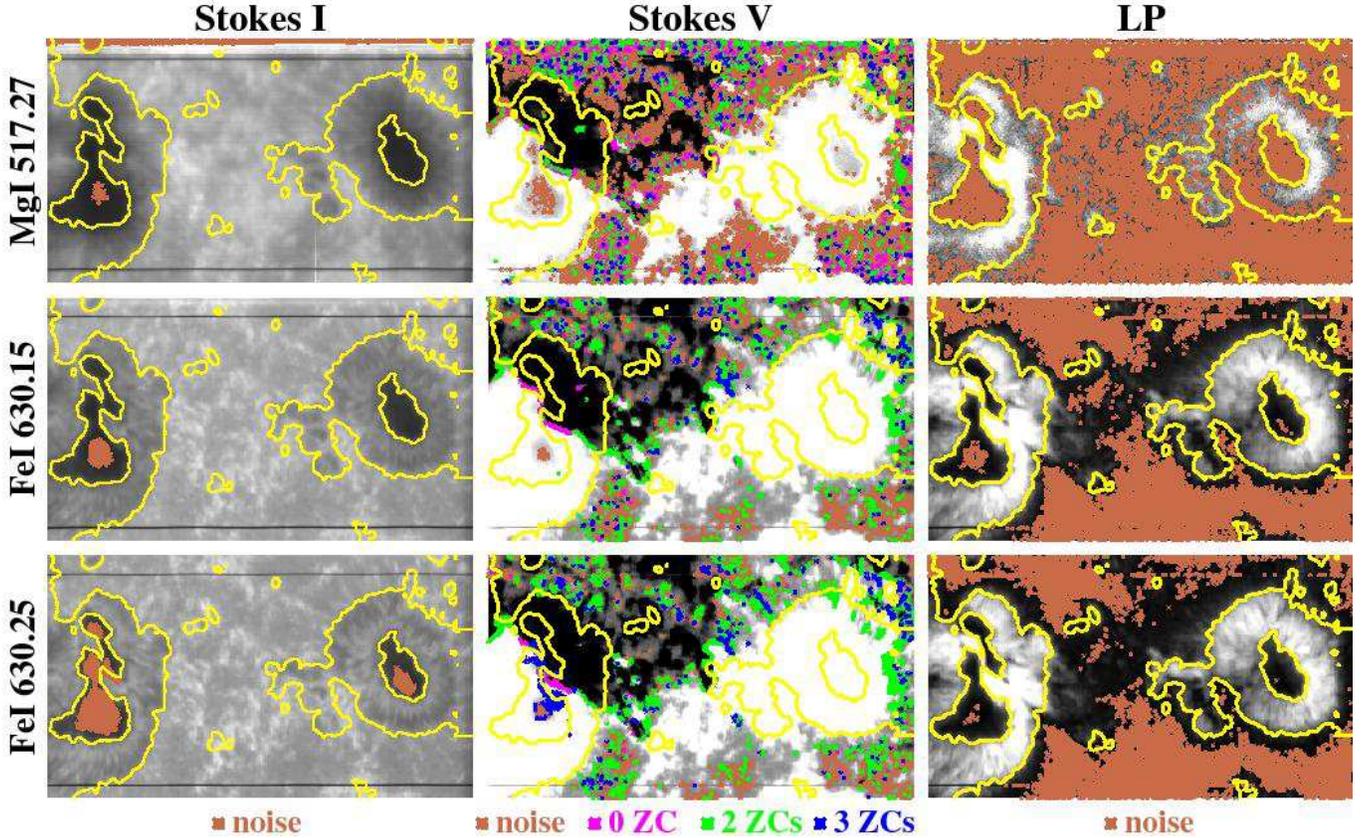}
  \caption{Stokes $I$ (line core intensity), Stokes $V$ (integration over Stokes $V$ blue lobe) and $LP$ (integration over $LP$ signal) images in the three spectral lines. The sunspot umbrae, penumbrae and pores, as identified on the basis of continuum intensity, are outlined by the yellow contours. The orange points on Stokes $I$, $V$ and $LP$ images indicate the locations where the Stokes $I$, $V$ and $LP$ profile has a S/N less than 7, respectively. The positions with abnormal Stokes $V$ profiles (red, green, and blue for zero, two, and three ZCs, respectively) are also denoted on each Stokes $V$ image accordingly. To enhance the visibility, the size of the colored points is larger than the real size of the corresponding image pixel.}
  \label{FIG:NZC}
\end{figure*}

\begin{figure*}[t]
  \epsscale{1.17}
  \plotone{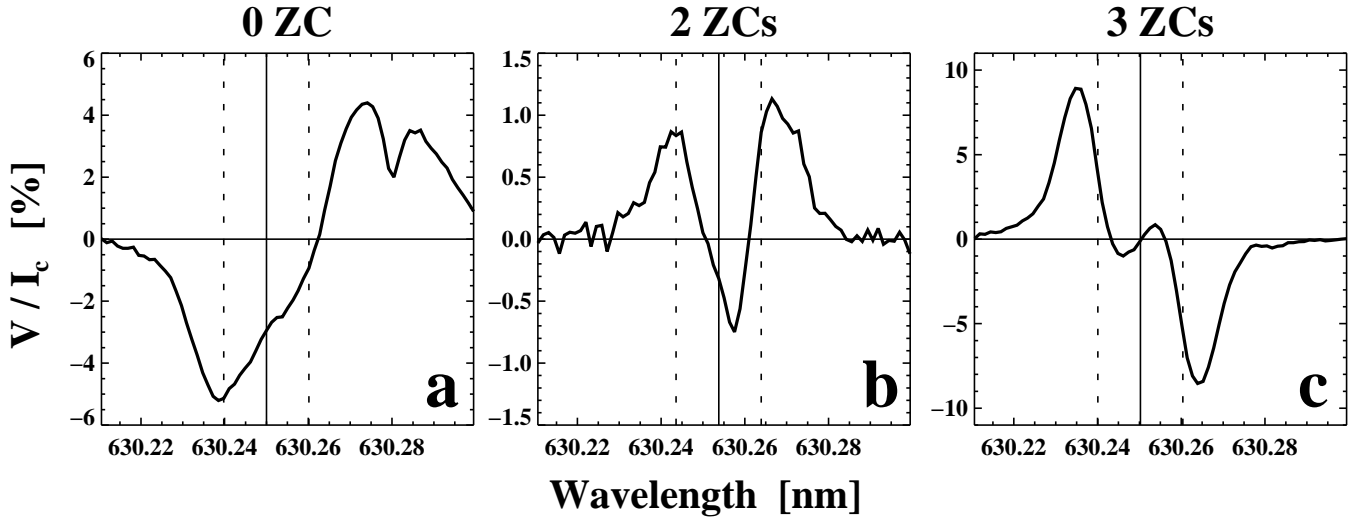}
  \caption{The sample abnormal Stokes $V$ profiles defined in this study with zero, two, and three ZCs within $\pm 5$~\kms\ from their Stokes $I$ line centers indicated by the vertical solid line. The vertical dotted lines mark the wavelength range corresponding to $\pm 5$~\kms\ from Stokes $I$ line center.}
  \label{FIG:ABVSample}
\end{figure*}

Figure~\ref{FIG:NZC} shows the Stokes $I$, $V$ and $LP$ images constructed by line core intensity and integrated Stokes $V$ and $LP$ signals, respectively. Images of the three spectral lines are arranged following the order of their formation heights (increasing height from bottom to top). The orange points are the locations of noisy profiles with S/N $<$ 7. The $LP$ spectra have more noisy profiles than Stokes $V$, because the Zeeman signals for $LP$ are essentially lower and near the disk center the transverse magnetic fields are weaker than the longitudinal fields for most of the areas in the low atmosphere. Compared to the photospheric \FeI\ spectra, the chromospheric \MgI\ polarimetric (Stokes $V$ and $LP$) spectra generally have lower S/N thus more useless profiles. The reasons are mainly due to the larger line width and the lower magnetic field strengths in the chromosphere. In the following, we first address the spatial distribution of abnormal Stokes $V$ profiles, which are important for evaluating the complexity of magnetic and flow fields \citep[e.g.,][and references therein]{Mickey1985SoPh...97..223M, Sigwarth2001}. We then present the results of Doppler shifts and asymmetric properties of the Stokes profiles, where all the noisy and abnormal profiles are excluded from analysis.

\begin{figure*}[t]
  \epsscale{1.17}
  \plotone{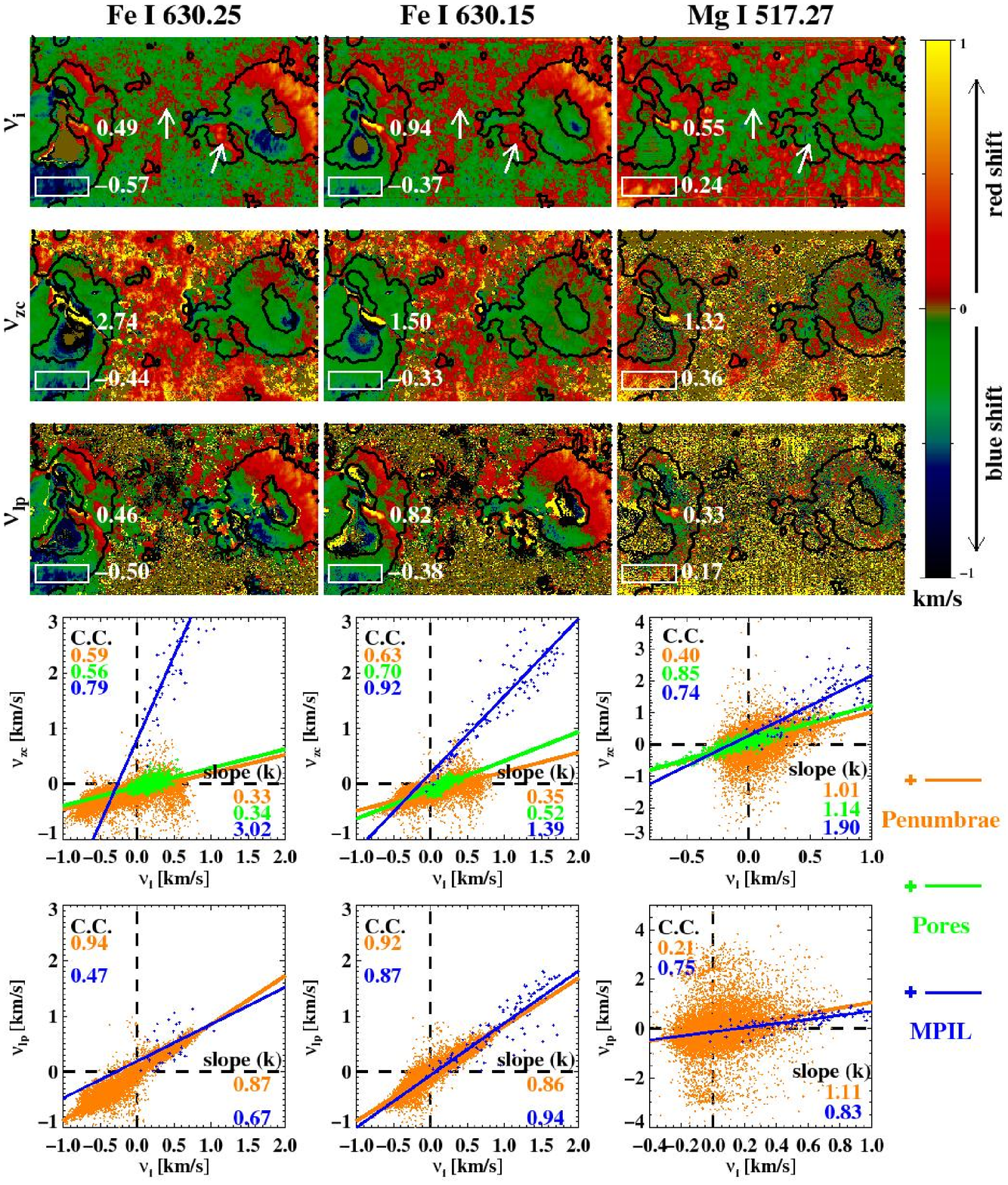}
  \caption{\textit{Upper panels}: Dopplergrams made using Stokes $I$ line center shifts ($\nu_i$), Stokes $V$ ZC shifts ($\nu_{zc}$) and $LP$ profiles shifts ($\nu_{lp}$) for the three spectral lines. The numbers in unit of \kms are the mean velocities within the strong and compact redshift region around the MPIL of the $\delta$ spot, and those in the penumbral region outlined by the white box. The spatial points with noisy and abnormal profiles are assigned a value of 0. Note that the methods of determining $\nu_i$, $\nu_{zc}$ and $\nu_{lp}$ may break down in some umbral areas for fully split lines. \textit{Lower panels}: Scatter plots of $\nu_i$ vs. $\nu_{zc}$ (\textit{forth row}) and $\nu_i$ vs. $\nu_{lp}$ (\textit{fifth row}) for each spectral line. The color-coded data points stand for different magnetic regions. The slope $k$ and the linear correlation coefficient (C.C.) from the line fits $\nu_{zc} = k\nu_i+b$ or $\nu_{lp} = k\nu_i+b$ for each group of data are labeled.}
  \label{FIG:VELO}
\end{figure*}

\subsection{Abnormal Stokes $V$ Profiles} \label{abnormal}
In the middle column of Figure~\ref{FIG:NZC}, locations with abnormal Stokes $V$ profiles are marked using colored points. They are divided into three kinds according to the number of ZC within a range of $\pm 5$~\kms\ from their Stokes $I$ line centers as shown in Figure~\ref{FIG:ABVSample}. Profiles with two ZCs ({\it green}) are most frequently seen, which have three lobes with the central one opposite to the other two (Fig.~\ref{FIG:ABVSample}$b$). They concentrate in the regions of magnetic polarity inversion line (MPIL) and the outer penumbral boundaries. They are mainly caused by the presence of mixed polarities associated with different velocities \citep[e.g.,][]{Sigwarth2001}, whereby the two polarities can be present beside each other (i.e., within one resolution element) or along the LOS over the line formation region \citep{SanchezAlmeida+Lites1992ApJ...398..359S, Solanki+Montavon1993A&A...275..283S}. In outer penumbral regions, the number of two ZCs decreases with height, which implies that the mixed polarity effect is weaker at higher atmospheric levels. This is probably due to the fact that the orientation of outer penumbral fields changes from horizontal or even downward to more vertical when they extend from the photosphere to the chromosphere \citep[][and references therein]{Choudhary+Bala2007}. However, in the central part of the MPIL of the $\delta$ spot there are more locations with two ZCs in the \MgI\ image. We speculate that this may point to a stronger mixed polarity effect around the highly non-potential MPIL in the chromosphere.

The central part of the $\delta$ spot's MPIL is also populated by locations without ZC ({\it red}). Such locations may involve magnetic components carrying high speed flows, which cause the ZC to shift far from the line center (see Fig.~\ref{FIG:ABVSample}$a$). Since the ZC position of Stokes $V$ is susceptible to the presence of velocity gradients through the line formation region \citep{LopezAriste2002ApJ...564..379L} or distortion of line profiles by noise, it is also possible that the profiles at these locations are highly asymmetric or distorted.

There exists another kind of profile with three ZCs ({\it blue}). They could be caused by residual noise or mixed polarity effect. Other than this, the three ZCs profiles within the umbra seen in 630.25~nm Stokes $V$ image are most likely due to the magneto-optical effect (see Fig.~\ref{FIG:ABVSample}$c$) caused by strong Zeeman effect \citep[e.g.,][]{West+Hagyard1983SoPh...88...51W}.

\subsection{Doppler Shifts of Stokes $I$, $V$ and $LP$ Profiles}
In the upper panels of Figure~\ref{FIG:VELO}, we show Dopplergrams ($\nu_i$,  $\nu_{zc}$ and $\nu_{lp}$) for each spectral line. They are constructed using the LOS velocities derived from the shifts of Stokes $I$, $V$ and $LP$ profiles at each spatial point, respectively. We note the followings: (1) For the two photospheric lines, the spatial distribution of penumbral Evershed flow in $\nu_i$ Dopplergrams is similar to that in $\nu_{lp}$ but significantly different from $\nu_{zc}$ Dopplergrams. (2) The variation of Evershed flow with height can be seen by the three spectral lines. Interestingly, the Evershed flow reverses its sign from the \FeI\ to \MgI\ Dopplergrams in some places as illustrated by a center-side penumbra ({\it white box}), which is known as the inverse Evershed effect observed in the chromosphere \citep[e.g.,][]{solanki2003}. While in some other places, the sign reversal has not yet happened or is rudimentary. This provides a qualitative evidence on the formation height of the \MgI\ b$_2$, which is at or just above the TMR. (3) The \FeI\ $\nu_i$ Dopplergrams show that small pores and plages lying in between the two sunspots are generally surrounded by red-shifted downflows, some of which do not appear in the \MgI\ $\nu_i$ Dopplergram as pointed by white arrows on the top three panels of Figure~\ref{FIG:VELO}. In contrast, the \FeI\ $\nu_{zc}$ Dopplergrams show ZC redshifts all over the plage regions. We suspect that the red-shifted $\nu_{zc}$ signal may not represent real down flows in flux tubes but could result from the combination of the susceptibleness of ZC position to the large Stokes $V$ asymmetry often observed by photospheric lines in plage regions and the limited spectral resolution due to smoothing \citep[see][]{Solanki+Stenflo1986A&A...170..120S, LopezAriste2002ApJ...564..379L}.

We quantitatively compare the $\nu_i$ with $\nu_{zc}$ and $\nu_{lp}$ in different magnetic regions for each spectral line by making scatter plots of $\nu_i$ vs. $\nu_{zc}$ and $\nu_i$ vs. $\nu_{lp}$ shown in the forth and fifth row of Figure~\ref{FIG:VELO}, respectively. Since the $LP$ signal is only strong enough in penumbrae and MPIL, velocities from these regions are plot in $\nu_i$ vs. $\nu_{lp}$ panels. Each group of data points are fit with a form of $\nu_{zc} = k\nu_i+b$ or $\nu_{lp} = k\nu_i+b$, and their linear correlation coefficients (C.C.) are also obtained. We discuss the scatter plots for different regions as follows.

\textit{Pores}---For the two photospheric \FeI\ lines, $|\nu_{zc}| < |\nu_i|$ (i.e., $k < 1$) is observed with moderate correlation (C.C.$=0.56$ and $0.70$, respectively). For the \MgI\ line, $|\nu_{zc}| \approx |\nu_i|$ (i.e., $k \approx 1$) with strong correlation (C.C.$=0.85$) is observed. It is well established that the magnetic filling factor (MFF) is almost 1 and the magnetic fields are nearly vertical in the pores. This is even true for pores in heights above the photosphere, which explains that $\nu_{zc}$ show large similarity with $\nu_i$ as seen by the \MgI\ line. While down to the photosphere near the edge of pores the field lines may be more inclined and the MFF may be less than 1 in such an observation with median spatial resolution \citep[e.g.,][]{leka+steiner2001}. Strong and narrow field-free down flows with speed larger than that within the pores have been found immediately around pores in the photosphere \citep{Sankarasubramanian+Rimmele2003ApJ...598..689S}, which gives rise to $|\nu_i|$ instead of $|\nu_{zc}|$. This might be responsible for the observed $|\nu_{zc}| < |\nu_i|$ in the \FeI\ lines.

\textit{Penumbrae}---The penumbral magnetic fields are known to consist of a more vertical component and a more horizontal component in the photosphere \citep[interlocking-comb structure, e.g.,][]{Ichimoto+etal2007PASJ...59S.593I}. For this region close to the disk center, $\nu_{zc}$ have more contribution from the more vertical component than from the other component. On the contrary, $\nu_{lp}$ have more contribution from the more horizontal component. Figure~\ref{FIG:VELO} shows that $\nu_i$ has much better spatial and magnitude correlation with $\nu_{lp}$ than with $\nu_{zc}$ for the two \FeI\ lines. Moreover, the slopes of $\nu_i$ vs. $\nu_{lp}$ (k$=0.87$ and $0.86$, respectively) are close to 1 and much larger than those of $\nu_i$ vs. $\nu_{zc}$ (k$=0.33$ and $0.35$, respectively). In other words, the $\nu_{lp}$ is very close to $\nu_i$ and their magnitude is much larger than $|\nu_{zc}|$. These results directly support that the penumbral Evershed flows are not only magnetized but mainly carried by the horizontal magnetic component as previously found by other authors using different observations and methods \citep[see, e.g.,][and references therein]{rimmele+marino2006, Ichimoto+etal2008A&A...481L...9I}. While for the \MgI\ line, $\nu_i$ is observed similar to both of $\nu_{lp}$ and $\nu_{zc}$ (i.e., $k \approx 1$) with small correlation coefficients (C.C.$=0.21$ and $0.40$, respectively). This implies that rising up to the TMR the penumbral magnetic and flow structures are different from those in the photosphere. The large scatter of the data points resulted from noisier and weaker polarimetric signal of the \MgI\ line hinders us from drawing solid conclusion. We thus speculate that the interlocking-comb structure of the penumbral magnetic and flow field might be a shallow phenomenon mainly in the photosphere.

\textit{MPIL}---For a compact elongated region around the MPIL and within the common penumbra of the $\delta$ spot, apart from noise dominating the \FeI\ 630.25 $\nu_i$ and $\nu_{lp}$ signal, the Dopplergrams show strong redshift with a mean speed conspicuously larger than that of simultaneously observed Evershed flows. $\nu_{zc}$ and $\nu_{lp}$ are generally well correlated with $\nu_i$. The linear fittings show that $|\nu_{lp}|$ and $|\nu_i|$ are close to each other and much smaller than $|\nu_{zc}|$. The large $|\nu_{zc}|$ may be artificial signal resulted from large Stokes $V$ asymmetry and limited spectral resolution as discussed above. It is also possible that the mixed polarities there contain a component possessing a high speed downflow, as observed earlier by \citet{MartinezPillet1994ApJ...425L.113M}. Detailed examination of the large redshift requires two-component Stokes inversion of individual profiles, which is out of the scope of this study.

\begin{figure*}[t]
  \epsscale{1.17}
  \plotone{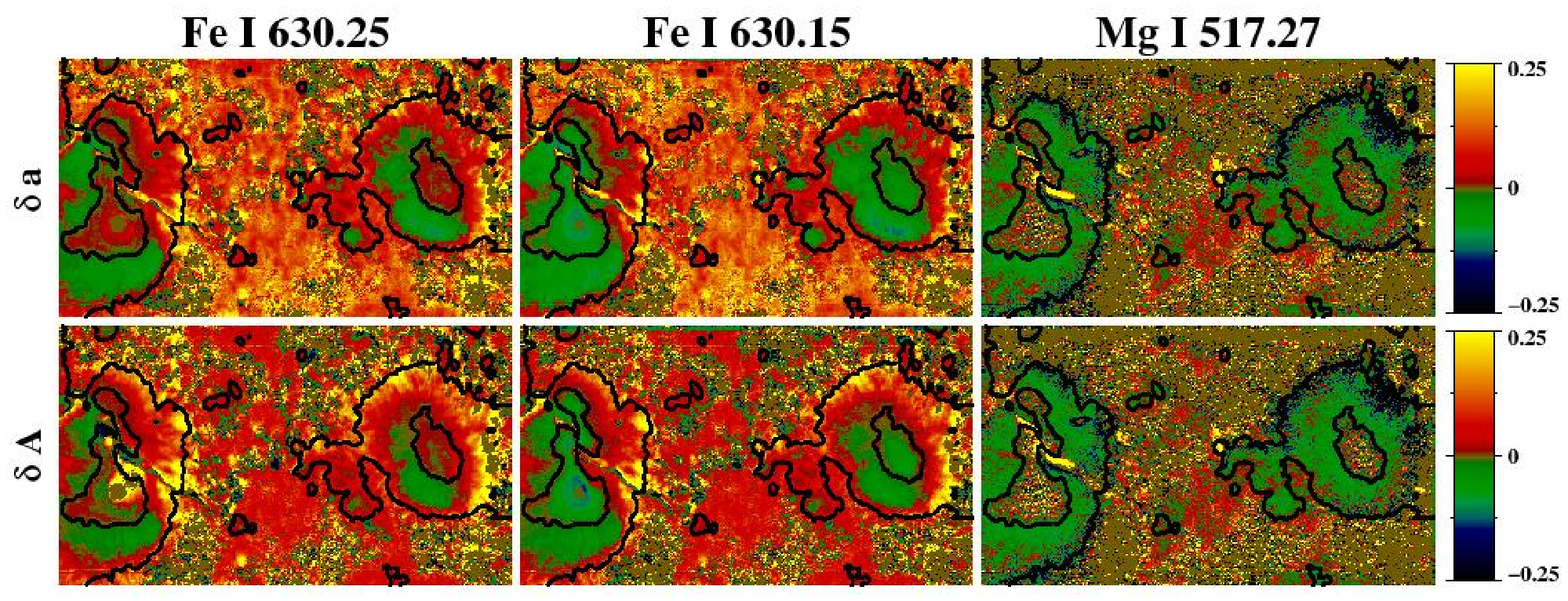}
  \caption{Images constructed with values of Stokes $V$ amplitude asymmetries ($\delta a$, first row) and area asymmetries ($\delta A$, second row) at each spatial point for the three spectral lines. The magnitude of asymmetry is indicated by the scales at right. The spatial points with noisy and abnormal profiles are assigned a value of 0.}
  \label{FIG:ASYMIMGS}
\end{figure*}

\begin{figure*}[t]
  \epsscale{1.17}
  \plotone{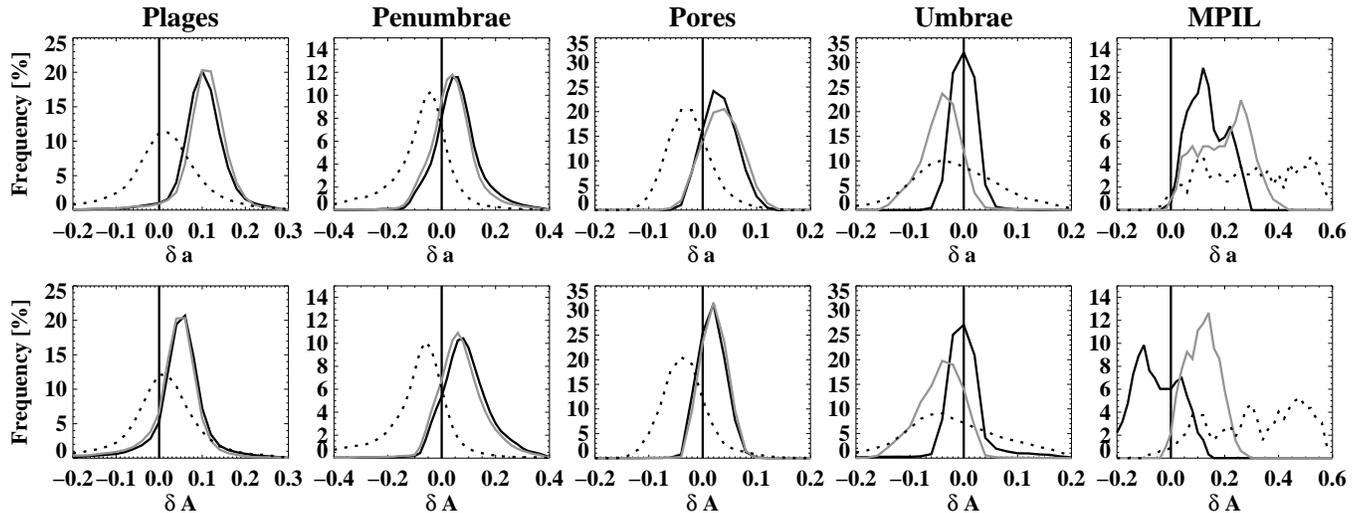}
  \caption{The distribution of Stokes $V$ amplitude asymmetries ($\delta a$, first row) and area asymmetries ($\delta A$, second row) in different magnetic regions for the three spectral lines. The black, gray, and dotted lines are for the \FeI\ 630.25, \FeI\ 630.15, and \MgI\ 517.27 observations, respectively.}
  \label{FIG:HISTOGRAM}
\end{figure*}

\begin{figure*}[t]
  \epsscale{1.17}
  \plotone{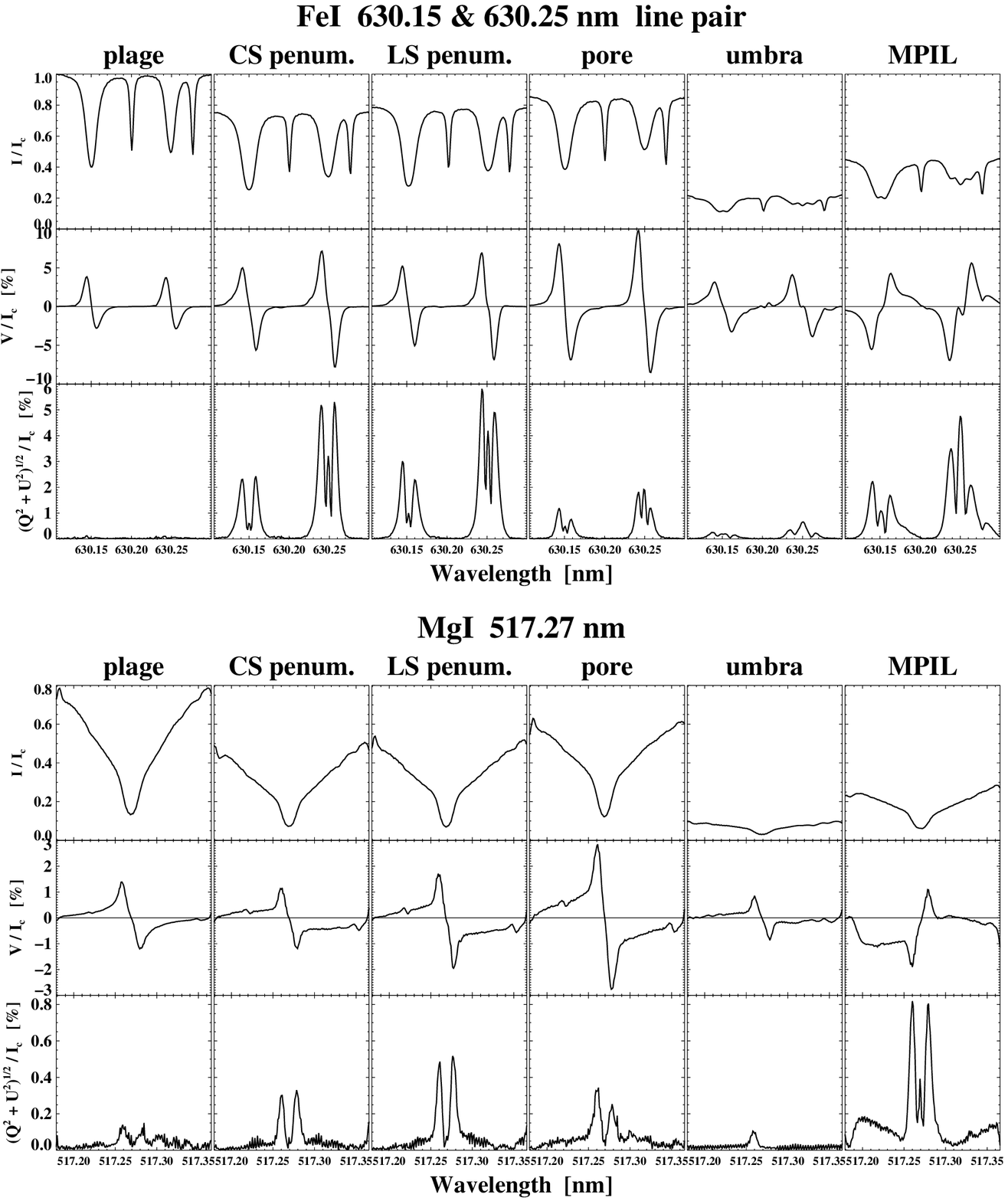}
  \caption{Typical Stokes $I$, $V$ and $LP$ profiles of the \FeI\ line pair and the \MgI\ b$_2$ line averaged over different magnetic regions: plage, center-side (CS) penumbra, limb-side (LS) penumbra, pore, umbra, and the compact strong red-shifted region around the MPIL of the $\delta$ spot.}
  \label{FIG:AVGPROF}
\end{figure*}

\subsection{Asymmetry of Stokes Profiles} \label{asymmetry}
Asymmetries of both Stokes $V$ and $LP$ profiles were calculated. We present results from Stokes $V$ only because the noisier and weaker $LP$ signals produce large uncertainty in determined asymmetries although their spatial distributions are qualitatively similar to those of Stokes $V$. Figure~\ref{FIG:ASYMIMGS} shows the spatial distribution and spectral variation of the Stokes $V$ amplitude asymmetry ($\delta a$) and area asymmetry ($\delta A$). We also plot their histograms for different magnetic regions in Figure~\ref{FIG:HISTOGRAM} in order to quantitatively compare the Stokes $V$ asymmetries among the three spectral lines. The typical Stokes $I, V$ and $LP$  profiles in those regions are also plotted in Figure~\ref{FIG:AVGPROF}. We again discuss various magnetic features separately as follows.

For plages, the red lobes are broader and shallower than the blue lobes for the two photospheric \FeI\ lines, which makes highly asymmetric Stokes $V$ profiles thus substantial positive $\delta a$ and $\delta A$ with $\delta a > \delta A$ generally, as previously found \citep{Stenflo+etal1984A&A...131..333S, Solanki+Stenflo1984A&A...140..185, Solanki+Stenflo1985A&A...148..123, Grossmann-Doerth+Keller+Schuessler1996A&A315.610, Sigwarth2001}. It is believed that the ``magnetic canopy'' effect and the small MFF in photospheric plages play important roles in producing such large positive asymmetries \citep[e.g.,][]{leka+steiner2001}. A noteworthy finding here is that the chromospheric \MgI\ Stokes $V$ profiles are much more symmetric as can be seen that the most probable $\delta a$ and $\delta A$ are very small positive values. \citet{Briand+Solanki1998A&A...330.1160B} also observed smaller Stokes $V$ asymmetry in the \MgI\ b$_2$ line compared to a photospheric line in network regions. It was found that for small or weak magnetic structures, such as plage and network, the degree of Stokes asymmetries decreases if MFF increases \citep[e.g.,][]{Zayer+etal1990A&A...239..356Z, martinezpillet+lites+skumanich1997, Sigwarth+etal1999A&A349..941}. The dependence of amplitude of Stokes $V$ asymmetries on some line parameters, such as Doppler shift, Zeeman shift, line strength and width, has also been heuristically considered within the frame of a simplified magnetic canopy model \citep{grossmanndoerth+schuessler+solanki1989A&A...221..338G}. There the asymmetry is expected to be small for a same downdraft velocity if the line is broad and not strongly Zeeman split. Our result confirms these trends from observational point of view. To sum up, both the broader line width and the larger MFF in the formation layer of the \MgI\ b$_2$ line are responsible for the observed small asymmetry. Together with the aforementioned finding that some of the conspicuous downflows surrounding the plages disappear in the \MgI\ $\nu_i$ Dopplergram, we therefore suspect that the ``magnetic canopy'' of plages could be a shallow phenomenon mainly below the TMR, which however needs more solid observational evidence to justify.

For penumbrae (especially their outer parts) and pores, $\delta a$ and $\delta A$ are primarily positive and negative in the photosphere and in the low chromosphere, respectively. Previous studies have shown that penumbrae and pores share some common properties. They both present an interface with the quiet sun and are surrounded by vortex convective flows with similar patterns \citep[e.g,][]{Deng+etal2007ApJ...671.1013D}. This study hence adds to the evidence of another common property, i.e., Stokes $V$ asymmetries and their variation with height. Unlike photospheric lines that can be easily synthesized under LTE conditions, the physical origins of the asymmetries of chromospheric lines formed in Non-LTE atmosphere are far from fully studied. Here we can only present a speculation for the sign reversal of Stokes $V$ asymmetries based on a logical analog of chromospheric line asymmetries to the better known photospheric line asymmetries that are mainly caused by velocity gradients. The change in sign of Evershed effect for penumbrae may be responsible for some but not all areas, because the Stokes $V$ asymmetries reverse sign all over the outer penumbral borders while the inverse of Evershed flow only occurs in some penumbral sectors. We speculate that the sign reversal of Stokes $V$ asymmetries in both outer penumbrae and pores could indicate that the velocity in and around them may change drastically from the photosphere to the low chromosphere. Figure~\ref{FIG:VELO} provides clues to some of the changes because some downflows in and around pores do not hold in the \MgI\ $\nu_i$ Dopplergram instead more upflows can be seen (e.g., in the scatter plot of the \MgI\ line). Moreover, \citet{Sankarasubramanian+Rimmele2003ApJ...598..689S} observed pores in different wavelengths and found that velocities in the narrow downflow region around pores tend to decrease with height at first, and then change to upflow in the low chromosphere. Such tendency is also visible in the MHD simulation of pores \citep[see Fig. 3 of][]{leka+steiner2001}. The Stokes $V$ asymmetries change from positive for the \FeI\ lines to negative for the \MgI\ line, if indeed all line asymmetries have a velocity gradient origin, thus provides important clues on the 3D structure of the flow field in and around pores and outer penumbrae.

For umbrae, since they lack of mass flows and have MFF of $\sim$1 in the low photosphere, both of the most probable $\delta a$ and $\delta A$ of the \FeI\ 630.25~nm line are close to 0. The large Zeeman splitting of the line in strong field may also contribute to the observed small asymmetry \citep{grossmanndoerth+schuessler+solanki1989A&A...221..338G}. However, the distributions of $\delta a$ and $\delta A$ of the other two spectral lines show a clear shift toward negative. Besides the noise induced inaccuracy of the determined asymmetry due to low light level in the umbra, it is also possible that as height increases umbrae become less uniform and static compared with their structure in the low photosphere.

For the compact region with strong redshifts around the MPIL of the $\delta$ spot, both $\delta a$ and $\delta A$ have prominent values that increase with height. It seems from Figure~\ref{FIG:AVGPROF} that the red wings of the Stokes $I$, $V$, and $LP$ profiles of the two photospheric lines obviously have further redward extension. Meanwhile, the blue damping wings of the \MgI\ 517.27~nm line are surprisingly enhanced, which remains a puzzling issue. We conjecture that the observed large asymmetries could be due to the combination of the following two effects: the mixed polarities or flows, and the large gradients in velocity, magnetic field vector, or temperature near the MPIL region. Considering all the unusual properties observed in this region, we suggest that the magnetic and flow fields around the MPIL of the $\delta$ spot are very complex, which deserve a deeper investigation in the future.

\section{SUMMARY}\label{sec:summary}
We have presented a comprehensive study of the Doppler shifts and asymmetric properties of the Stokes profiles, the uniqueness of which is especially in the combination of magnetic sensitive spectral lines that form at different heights spanning from the photosphere to the low chromosphere. Our quantitative analysis benefits from the simultaneous and high resolution spectro-polarimetric observation that include a chromospheric line. We summarize the major findings and interpretations as follows.

\begin{enumerate}
\item $\nu_i$ is very close to $\nu_{lp}$ but significantly different from $\nu_{zc}$ in penumbral areas near disc center for the photospheric \FeI\ lines, which provides direct and strong evidence that the penumbral Evershed flows are magnetized and mainly carried by the horizontal magnetic component.

\item The rudimentary inverse Evershed effect observed by the \MgI\ b$_2$ line provides a qualitative evidence on its formation height that is around or just above the TMR.

\item $\nu_{zc}$ and $\nu_{lp}$ in penumbrae and $\nu_{zc}$ in pores generally approach their $\nu_i$ observed by the chromospheric \MgI\ line, which is not the case for the photospheric \FeI\ lines. We speculate that the interlocking-comb structure of the penumbral magnetic and flow field might be a shallow phenomenon mainly in the photosphere.

\item The Stokes $V$ asymmetries exhibit similar behavior in the regions of outer penumbrae and pores. Both $\delta a$ and $\delta A$ evolve from primarily positive values observed by photospheric \FeI\ lines to primarily negative values observed by the chromospheric \MgI\ line.

\item In plage regions where the ``magnetic canopy'' effect is present, Stokes $V$ profiles are highly asymmetric in the photosphere but more symmetric in the low chromosphere. Moreover, some of the photospheric downflows surrounding the plages disappear in the \MgI\ Dopplergram. These suggest that the plage ``magnetic canopy'' could be a shallow phenomenon mainly below the TMR.

\item All the $\nu_i$, $\nu_{zc}$ and $\nu_{lp}$ Dopplergrams reveal strong redshifts in a compact region around the MPIL and within the common penumbra of the $\delta$ spot. Large positive Stokes $V$ asymmetries are also present and even strengthen with height. These unusual phenomena are manifestations of the complex nature of the 3D topology of the highly non-potential region.
\end{enumerate}

We conclude that analysis of Stokes profiles using both the photospheric and chromospheric spectral lines can furnish crucial information on the 3D structure of the magnetic and flow fields. Similar study using other chromospheric lines will be essential to see if the behavior and trend of velocities and asymmetries observed by the \MgI\ $b_2$ line is general for all the chromospheric lines. The speculations made in this study, such as the velocity gradient origin of the chromospheric line asymmetries and the thinness of the interlocking-comb structure of penumbral magnetic and flow fields, need further investigation to justify. To advance firmly towards a sound and reliable measurement of magnetic fields in the chromosphere, not only more polarimetric observations using chromospheric lines are needed, but also a better understanding of their complicated profile formation conditions is critical.

\acknowledgments
We are truly grateful to Dr. S. Solanki for careful reading of the paper and many constructive suggestions throughout this study. Thanks are also given to Dr. C. Liu and A. Cookson for polishing and improving the manuscript and Drs. L. Bellot Rubio, K. E., Rangarajan, and H. Wang for helpful discussions. We thank the referee for valuable comments that help us to improve the paper. N.D. and D.P.C. were supported by NASA grant NNX08AQ32G and NSF grant ATM 05-48260. We acknowledge the use of the HAO/NSO Advanced Stokes Polarimeter. The National Solar Observatory is operated by the Association of Universities for Research in Astronomy under a cooperative agreement with the National Science Foundation, for the benefit of the astronomical community. This research has made use of NASA's Astrophysics Data System Bibliographic Services.

\end{document}